# Identifying defect-tolerant semiconductors with high minority carrier lifetimes: Beyond hybrid lead halide perovskites

Riley E. Brandt,[1] Vladan Stevanović,[2,3] David S. Ginley,[2] and Tonio Buonassisi[1]

[1]*Massachusetts Institute of Technology, 77 Massachusetts Avenue, Cambridge, Massachusetts, 02139, USA*

[2]*National Renewable Energy Laboratory, 15013 Denver West Parkway, Golden, Colorado, 80401, USA*

[3]*Colorado School of Mines, 1500 Illinois Street, Golden, Colorado, 80401, USA*

The emergence of methyl-ammonium lead halide (MAPbX$_3$) perovskites motivates the identification of unique properties giving rise to exceptional bulk transport properties, and identifying future materials with similar properties. Here, we propose that this "defect tolerance" emerges from fundamental electronic structure properties, including the orbital character of the conduction and valence band extrema, the effective masses, and the static dielectric constant. We use MaterialsProject.org searches and detailed electronic-structure calculations to demonstrate these properties in other materials than MAPbX$_3$. This framework of materials discovery may be applied more broadly, to accelerate discovery of new semiconductors based on emerging understanding of recent successes.

Corresponding authors:   Riley E. Brandt        rbrandt@alum.mit.edu

                         Tonio Buonassisi       buonassisi@mit.edu

**I. INTRODUCTION**

Many semiconductors have been studied over the last century for their possible use in photovoltaics (PVs), light-emitting diodes (LEDs), computing devices, sensors, and detectors. Of these, only a select few have achieved sufficient optoelectronic performance to transition into industrial manufacturing, and their identification and development has been slow and mostly empirical. Recently, high-throughput computation and a deeper understanding of the physics-based requirements for high performance have created the potential for an accelerated identification of functional materials with manufacturing potential. For example, materials screening criteria have been proposed to better focus the search for novel candidate PV materials; they include the optical bandgap energy[1,2] and abruptness of absorption onset,[3,4] in addition to boundary conditions of elemental abundance[5] and manufacturing cost.[6,7] However, many materials have met these criteria and yet have not achieved industrially relevant conversion efficiencies (in excess of 10–15%) due to low minority carrier lifetimes or diffusion lengths, *e.g.*, in the case of $Cu_2ZnSn(S,Se)_4$,[8] SnS,[9] and others.[10] Meanwhile, PV devices have emerged based on methyl-ammonium lead iodide ($MAPbI_3$) and closely related halides (herein referred to as $MAPbX_3$). $MAPbI_3$ is a semiconductor which has demonstrated exceptional minority carrier lifetimes of 280 ns (in the mixed iodide-chloride composition)[11] and diffusion lengths up to 175 microns,[12] comparable to the best single-crystal semiconductors. This, in addition to meeting the criteria above, has resulted in a dramatic realization of photovoltaic conversion efficiencies up to 20.1%[13,14] in 2015, from around <4% in 2009.[15]

This paper examines whether the dramatic success of $MAPbX_3$ in PVs can be used as a basis to expand design criteria to identify new potential high-performance optoelectronic materials. One of the most compelling questions engendered by $MAPbX_3$ as an optoelectronic material is the degree to which it is unique, and whether its success can lead to the identification of materials with improved stability and lower toxicity, yet similar high performance. Clearly, as in previous design criteria the high optical absorption coefficient is important, but also essential are the long carrier diffusion lengths observed in $MAPbX_3$, enabled by high minority-carrier lifetime ($\tau$) and mobility ($\mu$).[11,16,17] The importance of $\tau$ and $\mu$ for device performance has been established[18] for the most highly performing PV materials, including silicon,[19] cadmium telluride,[20,21] copper indium gallium diselenide,[22] and gallium arsenide.[23]

Oddly, these more direct transport parameters, $\tau$ and $\mu$, are not traditionally considered essential screening criteria for novel candidate PV materials. This may be partially a consequence of the difficulty in measuring and/or calculating these parameters. The direct measurement of minority carrier $\tau$ and $\mu$ requires ultrafast electronic or optical sensors to capture transients,[24–27] or strong steady-state signals,[28,29] and must be performed with PV-device-relevant illumination conditions, electrical fields, and transport directions. Calculating $\tau$ and $\mu$ from first principles is even more challenging, given the lack of well established and high-throughput methods to directly calculate electron-phonon interactions and/or trap capture cross sections. While it is possible to obtain some information about carrier mobility from effective masses, and maximum radiative lifetime from the band structure, these are known to be insufficient and can be misleading when point and structural defects limit these transport properties.[18]

In this paper, we use established recombination models to identify the underlying electronic structure parameters of MAPbX$_3$ that are likely to underpin the high experimentally observed $\tau$ and $\mu$, including: dielectric constant, effective mass, band bonding character, and band dispersion (*i.e.*, band extrema relative to vacancy levels). We illustrate how these parameters can be computationally accessible using established variants of density functional theory (DFT). By using the basic properties of MAPbX$_3$ to define a set of search criteria, we discuss the potential to identify other novel materials with "defect-tolerant" transport properties, starting with the database at MaterialsProject.org.[30] This may yield materials with similar performance, but enhanced stability and lower toxicity.

## II. REDEFINING SEARCH CRITERIA FOR DEFECT-TOLERANT SEMICONDUCTORS

This section discusses the emerging explanations for high performance in MAPbI$_3$ and closely related materials, focusing on the unique aspects that make their performance exceptional: specifically their long non-radiative minority carrier lifetime, even in the presence of defects. A "defect tolerant" material is expected to either (a) form relatively few intrinsic or structural defects under high-throughput, low-temperature processing conditions, and/or more importantly (b) the extrinsic, intrinsic, or structural defects that do form have a very minimal effect on $\mu$ and $\tau$. We discuss models of how a material may achieve this defect tolerance, and suggest by example that the underlying electronic structure properties may not be unique to MAPbX$_3$, but rather be shared in a broader class of compounds.

## II.A. EXPLANATIONS FOR HIGH PERFORMANCE IN METHYLAMMONIUM LEAD IODIDE AND RELATED MATERIALS

Many authors have proposed explanations for the success of MAPbX$_3$ as a photovoltaic material. These include its large absorption coefficient, long electron and hole diffusion lengths, low exciton binding energies, low effective masses and high mobilities, and the presence of only shallow defects in the bandgap.[11,16,17,31,32] The existence of only shallow defects, and the disperse valence band, have both been tied to the presence of filled Pb 6$s$ orbitals, deriving from the partial oxidation of Pb relative to its Pb$^{4+}$ oxidation state. This orbital character has been identified by several authors, and explains both the shallow binding energy of defects and the atypical dependence of bandgap on strain or temperature in MAPbI$_3$.[33–35] This model for explaining the defect-tolerance and shallow defects has direct analogy to those developed earlier for CuInSe$_2$[36] and Cu$_3$N,[37] where instead the Cu$^+$ cation contributes Cu 3$d$ states to the valence band.

In addition, device-level observations support the claims that MAPbI$_3$ has excellent transport properties, including long carrier diffusion lengths significantly exceeding 1 micron,[11] and recently, up to 175 microns in single crystals.[12,38] Low non-radiative minority carrier recombination rates are also supported by measurements of high photoluminescence quantum yield (PLQY),[39] long carrier lifetimes,[40] and the high open-circuit voltage ($V_{OC}$) demonstrated by many devices. These properties are made more exceptional by the way in which MAPbI$_3$ is processed: low temperature solution processing results in equal or better performance to vacuum synthesis, which runs contrary to the higher purity, higher temperature, and more equipment-intensive fabrication strategies used to obtain high performance in conventional crystalline semiconductor materials.

We substantiate these earlier reports by applying density functional theory (DFT) to MAPbI$_3$. In the model pioneered by Zhang, Lany, Zakutayev *et al*,[36,37] defect tolerance emerges from having bonding orbitals at the conduction band minimum (CBM), and antibonding orbitals at the valence band maximum (VBM). This band structure is markedly different than that in most semiconductors, where the valence band is composed predominantly of bonding states, while the conduction band is composed of wavefunctions with dominant antibonding character. The defect tolerance that emerges from such an electronic structure is easily described using atomic vacancies as examples. Namely, if predominantly antibonding orbitals occur at the VBM and bonding at the CBM, the dangling (broken) bonds that are formed upon creating the vacancy (of any kind) will likely appear as resonances inside the bands, leaving the bandgap free of deep states that could act as carrier traps. As shown in Figure 1 for the valence band of MAPbI$_3$, this is a consequence of the position of valence *atomic* orbitals relative to the bonding and antibonding bands formed due to interactions. Analogous arguments are also valid for other types of intrinsic defects, such as interstitials and/or antisites, as well as for structural defects such as grain boundaries, which consist of many dangling bonds. Extrinsic impurities could, however, still introduce deep traps inside the bandgap depending on the position of their atomic orbitals relative to the energy bands of the host material.

These situations, in particular the antibonding character of the VBM, can frequently be found in metal-nonmetal systems with partially oxidized cations such as binary group-III halides or group-IV chalcogenides in 1:1 stoichiometry (*e.g.* TlBr, SnTe). Due to the partial oxidation of Pb, the MAPbI$_3$ compound also exhibits similar electronic structure features.[41] As shown in Figure 1, there are three types of interactions that contribute to the valence band in MAPbI$_3$. The first results from the overlap of Pb(6$p$)-I(5$p$) atomic orbitals leading to the creation of the deeper portion of the valence band, which is predominantly of I(5$p$) character, and the bottom part of the conduction band, which is composed mostly of Pb(6$p$) atomic orbitals. This interaction is responsible for chemical bonding as it implies charge transfer from the Pb(6$p$) to the I(5$p$) orbitals. In addition, Pb(6$s$)-I(5$p$) interaction leads to the creation of two bands: a deeper valence band (bonding), and the antibonding maximum of the valence band. Finally, only one of the I(5$p$) orbitals is oriented along the Pb-I bond, which is a favorable direction to interact with Pb(6$s$) and Pb(6$p$). The other two, oriented orthogonally, can only interact with the 5$p$ orbitals from other iodine atoms forming the middle of the valence band as shown in Fig. 1.

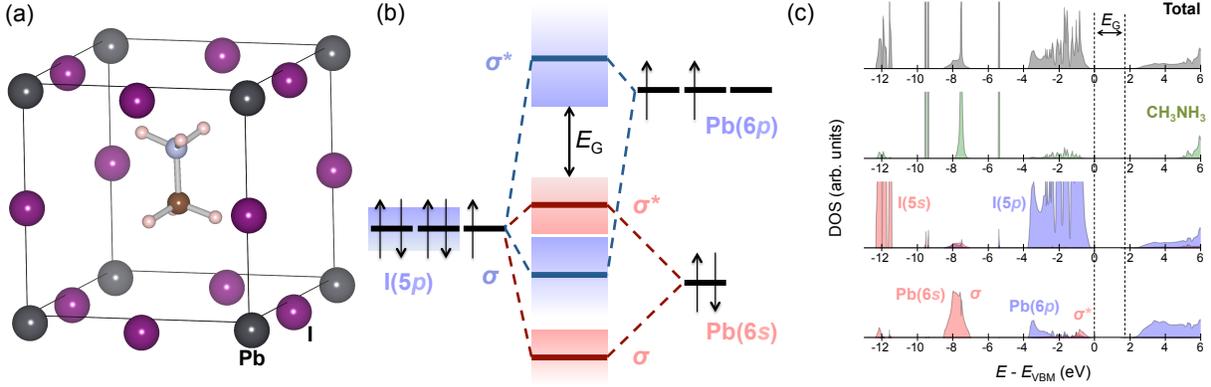

Fig. 1: (a) Crystal structure of cubic (high $T$) phase MAPbI$_3$, including the MA cation at the cage center; (b) Bonding/antibonding orbitals in MAPbI$_3$ demonstrating the formation of energy bands relative to isolated $p$ and $s$ atomic orbital energies; (c) Density of states (DOS) of MAPbI$_3$ separated into cation and anion $s$ and $p$ orbital contributions. The DOS is computed using density functional theory including the effects of spin-orbit interactions, with the band-gap adjusted to the value reported in Ref [42]. (1.85 eV computed using the many body GW method with spin-orbit included). The MA$^+$ cation does not introduce states at the band edge.

As a result of these interactions, the electronic bandgap is formed between the antibonding top of the valence band originating from the Pb(6$s$)-I(5$p$) interactions and the antibonding conduction band minimum resulting from the Pb(6$p$)-I(5$p$) interactions. Bonding-type defect-tolerance arguments can therefore only be applied to the defects that would tend to create valence band-derived states such as cation (Pb or MA) vacancies. On the other hand, if anion (iodine) vacancies form, the resulting dangling bonds will appear as resonances inside the conduction band only if the bandwidth and the dispersion of the conduction band are such that the CBM falls well below the energy of the Pb(6$p$) atomic orbitals. In the case of Pb, contrary to Sn and Ge compounds for example, this condition is more likely due to relativistic[43] spin-orbit interaction effects that increase the width of the Pb(6$p$) conduction band by ~2 eV based on our calculations. Thus, the fact that vacancy-type defects in MAPbI$_3$ are resonant in the bands is attributed to: (a) the presence of antibonding orbitals within the valence band that push the VBM energy up, and (b) relativistic effects which push the CBM energy down.

In summary, an emerging conclusion from both experimental measurements and first principles calculations is that MAPbX$_3$ benefits from an inherently defect-tolerant electronic structure.

**II.B. MODELING ELECTRONIC TRANSPORT IN THE PRESENCE OF DEFECTS**

To provide physical context for the claims above, and to translate them into electronic structure design criteria, we focus on transport models in defective materials: in particular, models for defect-assisted $\mu$ and $\tau$. These are semiconductor models (distinct from the Langevin recombination model, which is inappropriate for MAPbI$_3$).[40]

Defect-limited, monomolecular, or trap-assisted non-radiative recombination, $U_{\text{SRH}}$, is modeled by Shockley-Read-Hall (SRH) recombination statistics:

$$U_{\text{SRH}} = \frac{np - n_i^2}{\tau_{0,h}(n+n_1) + \tau_{0,e}(p+p_1)}, \qquad \text{Eq. 1}$$

where $n$, $p$, and $n_i$ are the electron, hole, and intrinsic carrier concentrations, respectively.[44] The low-injection electron and hole lifetimes $\tau_{0,e}$ and $\tau_{0,h}$, which we seek to maximize, are:

$$\tau_{0,h} = \frac{1}{N_t \sigma_{t,h} v_{\text{th},h}} \qquad \text{and} \qquad \tau_{0,e} = \frac{1}{N_t \sigma_{t,e} v_{\text{th},e}}, \qquad \text{Eq. 2}$$

where $N_t$ is the trap density, $\sigma_t$ is the capture cross section for electrons or holes, and $v_{\text{th}}$ is the drift velocity of carriers. At low injection conditions, the hole and electron lifetimes tend toward $\tau_h = \tau_{0,h}$ and $\tau_e = \tau_{0,e}$, respectively; at high injection conditions, both hole and electron lifetimes tend toward $\tau_h = \tau_e = \tau_{0,h} + \tau_{0,e}$. The terms $n_1$ and $p_1$ reflect the occupancy of trap states, where $E_t$ is the energy level of the trap state and $E_i$ the Fermi level in the intrinsic material (near mid-gap):

$$n_1 = n_i exp\left(\frac{E_t - E_i}{kT}\right) \qquad \text{and} \qquad p_1 = n_i exp\left(\frac{E_t - E_i}{kT}\right). \qquad \text{Eq. 3}$$

From this model, three conditions are necessary to limit SRH (trap-assisted) recombination: defect energy levels far from mid-gap, as these exponentially impact lifetime; low trap density $N_t$; and low capture cross-sections $\sigma$. Operating in high-injection conditions is helpful (where photogenerated carrier concentrations greatly exceed background doping densities), especially when strongly injection-dependent defects limit lifetime. $E_t$ and $N_t$ may be calculated from first principles, however this calculation is computationally expensive, so we seek more fundamental electronic properties accessible by high-throughput screening.

*The first, and most important property is the relative static permittivity, or dielectric constant $\varepsilon_r$.* A higher dielectric constant indicates a greater ability to screen charge, and may arise from electronic or ionic contributions, as well as the methylammonium molecular dipoles in the lattice.[45]

This screening means that the capture cross-section of a charged defect will be influenced by the dielectric constant. Assuming a Coulombic model for capture cross section, a capture event will occur when the electrostatic potential energy from a charged defect exceeds the thermal energy, $kT$, which occurs at a specific radius,[46] therefore:

$$\sigma_t = \frac{q}{16\pi(\varepsilon_r\varepsilon_0 kT)^2} \,, \quad \text{Eq. 4}$$

where $q$ is the elementary charge, and $\varepsilon_0$ is the vacuum permittivity. MAPbI$_3$ has demonstrated a large static dielectric constant of 60–70 or higher,[16,47,48] due to electronic, ionic, and molecular dipole contributions from the organic cation.[49] This charge screening is very beneficial, given the inverse square dependence of carrier capture cross-section, and in turn recombination rate, on the dielectric constant.

In addition, for a simple "hydrogenic" defect under Effective Mass Theory, the energy level of the defect sits deeper in the gap for lower dielectric constants,[44] *i.e.* the binding energy of a donor relative to the conduction band, $E_C$ is:

$$E_t - E_C = \frac{m_e^*}{\varepsilon_r^2}\frac{q^4}{2(4\pi\varepsilon_0\hbar)^2} \,, \quad \text{Eq. 5}$$

where $m^*_e$ is the effective mass of an electron, and $\hbar$ is the reduced Planck constant. Similar charge screening behavior has been used to explain the low binding energy of excitons in MAPbI$_3$[45] and shallow effective mass-like states.[50]

More important, however, is the energy level of non effective-mass like traps formed by vacancies and structural defects. As discussed above, band structures containing antibonding orbitals in the VBM, bonding in the CBM, and/or high dispersion within a band can lead to shallower defects, as defect energy levels are resonant with the conduction or valence bands. A second model proposed to explain the lack of deep trap states is that MAPbI$_3$ is more ionic than other semiconductors,[50] resulting in shallower or fewer defect states associated with dangling bonds. Analogous arguments have been made to explain defect tolerance in GaN[51] and (Cd,Zn)Te.[52]

*The second critical transport property is the minority-carrier mobility*, which is influenced by the effective masses of free carriers and by the frequency of scattering events. The lower the effective mass, or the more disperse the band edges, the higher the mobility. However, in typical polycrystalline materials at room temperature, the mobility will not reach this intrinsic limit due to several forms of scattering.

One way in which the mobility will be limited is by charged defect scattering, either at ionized impurities or grain boundaries. Again, a high dielectric constant reduces the spatial extent and the potential barrier of a charged defect, lowering its propensity to scatter free carriers. For ionized impurities, the mobility $\mu_{H,ii}$ is given by:

$$\mu_{H,ii} = r_{H,ii}\frac{128\sqrt{2\pi}(\varepsilon_r\varepsilon_0)^2(kT)^{3/2}}{\sqrt{m^*}N_t Z^2 e^3}\left[\ln(1+b) - \frac{b}{1+b}\right]^{-1}, \quad \text{Eq. 6}$$

where $r_{H,ii}$ is the Hall coefficient for ionized impurity scattering, $Z$ is the charge on the ionized site, $N_t$ is the trap density, $b = 24m^*kT/\hbar^2\beta_S^2$, and $\beta_S$ is the inverse screening length.[53] Thus, to reduce the impact of ionized impurities on mobility, it is again ideal to have a high dielectric constant, low effective mass, and low trap density. Grain boundary scattering is also reduced for larger dielectric constants, and when the defect levels associated with grain boundaries become shallower.[54]

Alternatively, the mobility limit may come from phonon scattering. The phonon-limited mobility is generally smaller in softer materials, given their higher concentration of phonons at a given temperature – the bulk modulus, $B$, and phonon-limited mobility $\mu_{ph}$ are correlated,[55] where $A_0$ is a constant:

$$\mu_{ph} = A_0 B (m^*)^{-2.5} \quad . \qquad \text{Eq. 7}$$

MAPbI$_3$ is a relatively soft material, with a bulk modulus predicted to be 22.2 GPa for the cubic phase.[56] This may explain why its mobility, measured around 10 cm$^2$/V/s,[40] is not large relative to III-V and group IV semiconductors. Claims of "high" mobilities may be relative to dyes and organic absorber materials from which much of the perovskite PV work derived. Thus, similar to the absorption coefficient of MAPbI$_3$, its mobility is necessary but insufficient to explain its high performance.

To compare these proposed performance criteria for PV, the ratio of the diffusion length to the absorption length in an absorber material must be maximized in order to produce high photocurrent in a PV device, as in this non-dimensional figure of merit[57]:

$$\nu = \frac{k_B T}{q} \mu \tau \alpha^2 \qquad \text{Eq. 8}$$

While historical PV screening approaches have focused on maximizing $\alpha$; here instead we focus on $\tau$. Given that the mobility and absorption coefficient are good, but not exceptional for MAPbI$_3$, we suggest that it is the minority carrier lifetime of MAPbI$_3$ that is most strongly responsible for its exceptional performance.

*Finally, it is worth mentioning that beyond point and structural defects, a disordered crystal structure can limit the conversion efficiency by introducing defect states near the band edges (Urbach tail).* MAPbI$_3$ demonstrates a very small Urbach tail energy.[58] This may be due to its low melting temperature, such that the material develops more crystalline order at lower processing temperatures (high homologous temperatures).[16]

To review, we summarize the priorities for high-performance PV absorbers in Fig. 2. While prior work has focused on screening for optical properties, we focus specifically on electronic structure properties that have a direct impact on transport, highlighted in bold.

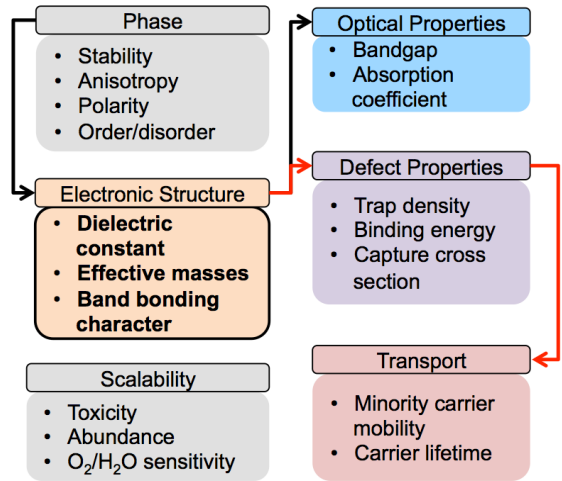

Fig. 2: Screening criteria for high-performance PV materials, focusing on the properties that enable excellent bulk transport in the presence of defects. Optical properties and scalability concerns have previously been emphasized in screening, and are important but insufficient to achieve high performance.

## II.C. TOWARD IDEAL ELECTRONIC STRUCTURE: THE IMPORTANCE OF PARTIALLY OXIDIZED, HEAVY POST TRANSITION METALS

Materials that contain a lone $6s^2$ or $5s^2$ pair of electrons (that do not participate in bonding) have the potential to share the high dielectric constant, low effective masses, and VB antibonding character that lead to defect tolerant transport properties. As a broad class, these are compounds formed from partially oxidized post-transition metals including In, Sn, Sb, Tl, Pb, and Bi. In the case of MAPbI$_3$, the partially oxidized Pb atom in its Pb$^{2+}$ charge state lends these beneficial properties – a large, polarizable cation leads to high Born effective charge and therefore large dielectric constants,[35] as well as a conduction band with greater bandwidth due to spin-orbit coupling, while filled $6s$ orbitals lead to a disperse valence band (low effective mass)[32] and antibonding orbital character in the valence band maximum.

Other partially oxidized post-transition metals including In$^+$, Sn$^{2+}$, Sb$^{3+}$, Tl$^+$, Pb$^{2+}$, and Bi$^{3+}$ share these properties, and are present in a wide variety of materials. This $N$-2 oxidation state is more commonly observed for compounds of Tl, Pb, and Bi, due to relativistic effects leading to tighter binding of the remaining $s$ electrons; however, lighter cations may also be stabilized in these oxidation states. Given their many beneficial properties, these cations appear increasingly in many optoelectronic applications: PbS and lead halide perovskites in PV; TlBr, PbI$_2$, and BiI$_3$ in X-ray detectors and scintillators;[59] Sn$^{2+}$-based TCOs;[60] and Pb and Bi-based ferroelectrics, photoferroelectrics,[61] piezoelectrics,[62] and thermoelectrics. For the thallous halides, analogous $6s^26p^0$ orbital explanations for low effective mass, antibonding VBM states, and a high dielectric constant were offered several years before similar work on MAPbX$_3$.[63,64]

To identify promising materials in this class, we developed a script to search through all materials in the Materials Project.org (MP) database.[30] This is implemented using the Materials Project API and the pymatgen library, to access information on calculated crystal and electronic structure for all compounds in the database.[65] This search examines approximately 27,000 non-metallic inorganic compounds with full band structures calculated, and extracts information on crystallographic point group, stability, band structure, and in particular the fractional density of states by element and orbital. To screen for the properties described above, we calculate the *s*-orbital fraction in the top 1 eV of the valence band density of states. This metric is a rough filter, and may mistakenly identify materials with band-inversion, as well as many metal hydrides and Au$^-$ compounds, wherein the anion species contributes a filled *s* orbital at the valence band edge.

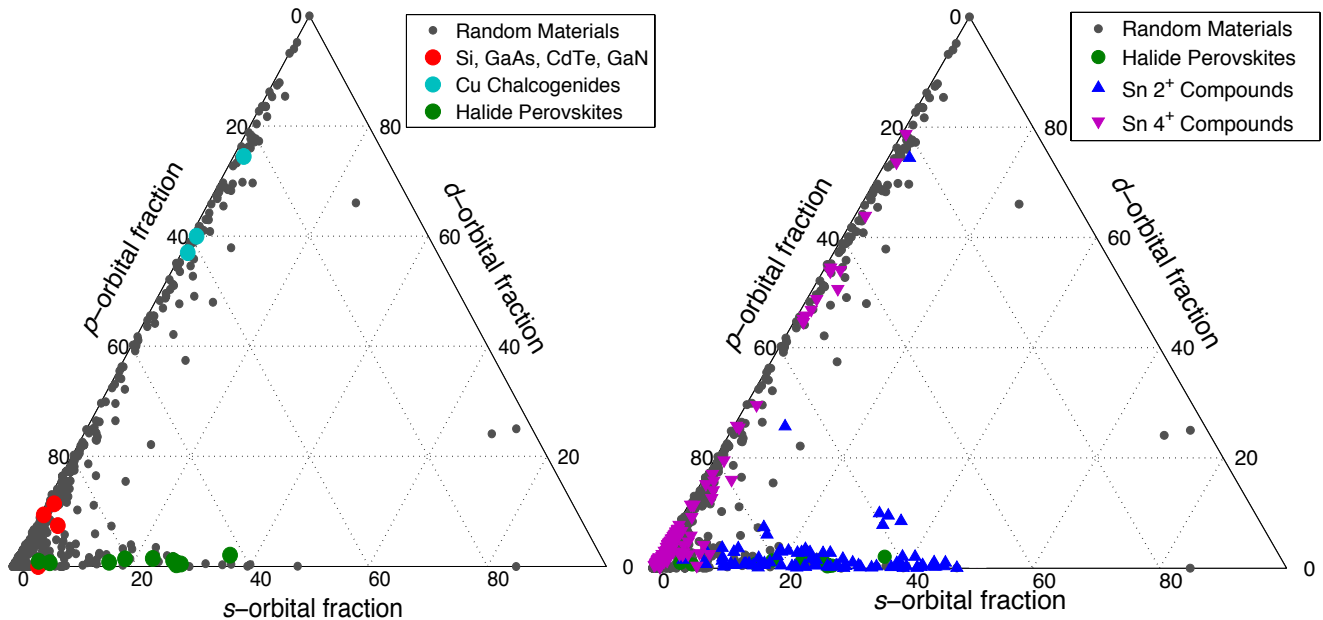

Fig. 3: Ternary plots demonstrating the partial density of states of the valence band maximum by orbital type, for a wide range of randomly selected materials. Overlayed on the left are the conventional zincblende and wurtzite semiconductors for PV, as well as Cu-chalcogenides showing higher *d*-orbital fraction, and the halide perovskites with higher *s*-orbital fraction. On the right, Sn-containing compounds demonstrate a distinct split along two axes, with Sn$^{2+}$ compounds demonstrating similar character to the halide perovskites.

In Fig. 3, we compare a random subset of the materials identified in this broad search, and note the stark differences in VBM density of states character for different classes of materials. Conventional tetrahedrally coordinated III-V and II-VI compounds all demonstrate almost exclusive *p* orbital character at the valence band maximum, as do the majority of compounds. Other defect tolerant Cu-chalcogenide semiconductors, such as CuInSe$_2$, demonstrate a large *d*-orbital character in the valence band due to full 3*d* cation orbitals. Meanwhile, the inorganic I-II-VII$_3$ family of perovskites sit along the bottom axis, reflecting their larger *s*-orbital fraction in the valence band density of states. However, they are not unique in this property. In addition, we plot a random fraction (30%) of the Materials Project Sn-cation compounds on these axes as well, separating them by those with $Sn^{2+}$ and $Sn^{4+}$ cations. The $Sn^{2+}$ cation-compounds cluster in the same region as the halide perovskites, while $Sn^{4+}$ compounds resemble the vast majority of semiconductors with anion *p*-orbital or transition metal cation *d*-orbital character.

Finding good candidate semiconductors with the *N*-2 oxidation state is not straightforward, for several reasons. Firstly, the use of more electronegative anions (oxygen, fluorine, and chlorine) tends to fully oxidize these cations. By pairing them with sulfur, selenium, iodine, and bromine, or by including a less electronegative cation such as an alkali metal, it is possible to stabilize the lower oxidation state. Secondly, the lone *s* orbitals are often stereochemically active,[66] leading to lower symmetry point groups, and a higher likelihood of indirect gaps. For example, for Sn compounds from the Materials Project database, 29% of those with $Sn^{2+}$ are found to have indirect bandgaps, while for $Sn^{4+}$, 47% have direct bandgaps. Anisotropic crystal structures may also lead to anisotropic transport properties and surface potentials,[67] which may adversely influence device performance. Lastly, these cations may convert to the higher oxidation states inadvertently, when exposed to oxidizers such as oxygen or moisture. In $CsSnI_3$ and $MASnI_3$, this $Sn^{4+}$ formation results in metallic behavior with very high carrier concentrations.[34,68]

Lastly, we note that all of the materials identified in the MaterialsProject.org search are inorganic, and we recognize the important role that the asymmetric, molecular cations such as methylammonium play in achieving large dielectric constants in these materials. Converting the inorganic materials identified herein into hybrid organic-inorganic analogues could lead to improved defect tolerance, but the presence of MA or other molecular cations is not the most important criteria in achieving defect-tolerance. Thus, replacement of Pb in $MAPbX_3$ with transition metal cations may not yield high-performing semiconductors.

## III. OUTLOOK: IDENTIFYING OTHER PROMISING MATERIALS CLASSES WITH EXCELLENT TRANSPORT PROPERTIES

By filtering the materials data drawn from the Materials Project database, we focus on several interesting classes that may share the beneficial properties identified above. Our search first identifies the class of halide perovskites ($CsPbI_3$, $CsSnI_3$, $RbPbBr_3$, *etc.*); to our knowledge prior optoelectronic screening efforts have not identified these as promising materials.

In the list below, we exclude $Pb^{2+}$ and $Tl^+$-containing compounds based on toxicity; there are many promising optoelectronic materials containing these atoms. Unfortunately, the beneficial properties of $Pb^{2+}$ and $Tl^+$ are also connected to their toxicity, as they mimic the large ionic sizes of $Ca^{2+}$ and $K^+$ in the human body. This excludes the binary lead and thallium halides and chalcogenides such as TlBr, as well as a number of ternary chalcohalides or chalcogenides such as $TlBiS_2$.

In addition, this search is limited to materials existing in the MaterialsProject.org database. New materials may be discovered or designed using the present framework, for example the recently discovered $Bi_2FeCrO_6$ perovskite,[69] which achieves an unusually small bandgap for a complex oxide ferroelectric and promising photovoltaic performance.

We compute the electronic structure from first principles for several classes of materials, which were identified through the broad computational survey. These classes are described below:

**1.) Binary chalcogenides and halides** ($BiI_3$, $Bi_2S_3$, $SbI_3$, $Sb_2S_3$, $SnI_2$, InI, *etc*.):

There are many binary iodides, sulfides, and selenides formed with $ns^2np^0$ cations, several of which have bandgaps in the range of interest for PV. Those with bandgaps 1.7 eV and greater, such as $BiI_3$, $PbI_2$, and TlBr, have been demonstrated successfully in X-ray detecting and scintillator applications, with high $\mu\tau$ products.[59,63,70]

**2.) Binary halides, sulfides, and selenides stabilized in cubic structures like NaCl and CsCl:**

Given the anisotropic transport properties that result from stereochemically active lone pairs in materials like SnS, more promising materials may result by stabilizing them in higher symmetry phases. Recently, it was demonstrated that SnS and SnSe may be stabilized in the NaCl structure by alloying with CaS.[71] Similarly, TlI may be stabilized in the CsCl phase by alloying,[72] suggesting a possible similar route with InI.

**3.) Chalcohalides** (BiOI, BiSI, BiSeI, SbSI, SbSeI, BiSBr, BiSeBr, *etc*.):

The $ns^2np^0$ chalcohalides (V-VI-VII compounds) range over visible to UV bandgaps,[73] and many demonstrate ferroelectric behavior at lower temperatures as well. Both Sb- and Bi-containing varieties have demonstrated photoconductive or photovoltaic effects.[74–77] SbSI and SbSeI have recently been identified for their potential applications as photoferroelectrics,[78] however those in the paraelectric phase may still demonstrate large Born effective charge, and a large static dielectric constant.

**4.) Ternary alkali chalcogenides** ($LiBiS_2$, $NaBiS_2$, $KBiS_2$, $RbBiS_2$, $CsBiS_2$, *etc*.):

Similar to the formation of I-III-VI$_2$ analogues of the zincblende II-VI compounds, one may form I-III-VI$_2$ rocksalt compounds based on the PbS and PbSe rocksalt phases.[79–81] These include Pb-free chalcogenides with an octahedrally-coordinated Bi$^{3+}$ cation.

**5.) Ternary halides:** (In$_3$SnI$_5$, InAlI$_4$, *etc.*):

In$^+$ and Tl$^+$ may be stabilized by several non-coordinating molecular anions such as AlI$_4^-$.[82] In addition, multiple cations can be stabilized in their lower oxidation states together, as in Sn$^{2+}$ and In$^+$ in the case of In$_3$SnI$_5$.

**6.) In+-II-VII$_3$ ternary halides** (CdInBr$_3$, CaInBr$_3$, *etc.*):

In$^+$ is also stabilized in a number of In-II-VII$_3$ compounds, which appear to have bandgaps outside of the range for PV, but which may exhibit similar defect tolerant properties.

**7.) Bismides and antimonides** (*e.g.*, KSnSb):

Sn$^{2+}$ and other partially-oxidized cations are stabilized in several bismides and antimonides, most of which have bandgaps too small for PV absorbers.

**8.) Octahedrally-coordinated metal halides** (Cs$_3$Bi$_2$I$_9$, Rb$_3$Bi$_2$I$_9$, K$_3$Bi$_2$I$_9$, Cs$_3$Sb$_2$I$_9$, *etc.*):

Bi$^{3+}$ and Sb$^{3+}$ form III$_2$X$_9^{3-}$ anions that consist of edge or face-sharing octahedra, bonding with alkali metals to form a variety of materials with different bandgaps, crystal structures, and ferroelectric phase transitions.[83–86]

**9.) Cs-containing compounds substituting 1+ molecular cations** (*e.g.*, (MA)$_3$Bi$_2$I$_9$, (FA)$_3$Bi$_2$I$_9$, *etc.*):

Lastly, we can identify a variety of Cs-containing materials, which could be converted to hybrid materials by the substitution of the Cs with an alkylammonium or other molecular cation.[85,87] A simple example is the family containing (MA)$_3$Bi$_2$I$_9$ or (FA)$_3$Bi$_2$I$_9$. A number of other molecular cations may be substituted as well based on size, analogous to the substitution explored in MAPbX$_3$,[88,89] to tune the bandgap or to form lower-dimensional structures.

As an illustration, we perform first principles (DFT) calculations on some of the above-listed materials to probe whether these materials have low effective masses and high static dielectric constants, in addition to containing cation antibonding orbitals in the VB, and compare these results to those of MAPbI$_3$. These DFT results are summarized in Table 1, including the bandgap ($E_G$), relative valence band ($E_V$) and conduction band energies ($E_C$), effective masses $m^*$ for holes and electrons from the DOS, the band degeneracy at the CBM and VBM extrema, and the ionic dielectric constant. DFT calculations are performed with spin-orbit coupling; we compare the band extrema energies with and without spin-orbit coupling to demonstrate the relativistic effects. DFT is known to consistently under-predict the bandgap, thus, we do not present the electronic component of the dielectric constant. Instead, we present only the ionic dielectric constant computed from density functional perturbation theory. In addition, the band gap values, both the minimum and the direct gap, are given to provide an indication of the type of optical transition at the lowest energy gap. The larger the difference between the minimal and the direct gap, the weaker the absorption onset is. More information on these calculations is supplied in the methods section.

**Table I**: DFT-computed band structures, band dispersion (with spin-orbit coupling, SO), effective masses, and ionic dielectric constants for several compounds of interest, as compared to the (MA)BX$_3$ family of perovskites.

| Formula | Space Group (International S.G. number) | $E_G$ [eV] (DFT+SO) | $E_G$ direct [eV] (DFT+SO) | $E_G$ (DFT+SO) − $E_G$ (DFT) | $E_V$ (DFT+SO) − $E_V$ (DFT) | $E_C$ (DFT+SO) − $E_C$ (DFT) | $m^*_{h,\text{DOS}}$ | $m^*_{e,\text{DOS}}$ | VBM band degeneracy | CBM band degeneracy | Ionic dielectric constant |
|---|---|---|---|---|---|---|---|---|---|---|---|
| InI | Cmcm (63) | 1.29 | 1.29 | -0.09 | 0.08 | -0.01 | 0.25 | 0.15 | 2 | 2 | 33.51 |
| SnI$_2$ | C12/m1 (12) | 1.58 | 1.74 | -0.18 | 0.10 | -0.08 | 1.49 | 0.35 | 4 | 2 | 28.84 |
| SbI$_3$ | R$\bar{3}$H (148) | 1.88 | 1.96 | -0.26 | 0.18 | -0.08 | 9.57 | 2.12 | 1 | 1 | 8.32 |
| SbI$_3$ | P12$_1$/c1 (14) | 1.86 | 1.93 | -0.25 | 0.18 | -0.07 | 8.88 | 1.72 | 2 | 1 | 8.43 |
| BiI$_3$ | R$\bar{3}$H (148) | 1.73 | 1.82 | -0.78 | 0.23 | -0.55 | 10.39 | 1.85 | 1 | 1 | 5.70 |
| BiI$_3$ | P$\bar{3}$1m (162) | 1.66 | 1.72 | -0.81 | 0.23 | -0.57 | 9.34 | 0.79 | 1 | 2 | 5.78 |
| Sb$_2$S$_3$ | Pnma (62) | 1.28 | 1.30 | -0.02 | -0.01 | -0.03 | 3.89 | 0.88 | 2 | 3 | 44.94 |
| Bi$_2$S$_3$ | Pnma (62) | 1.14 | 1.14 | -0.24 | -0.17 | -0.41 | 2.86 | 0.49 | 2 | 1 | 400.33 |
| BiOI | P4/nmmS (129) | 1.38 | 1.49 | -0.13 | 0.00 | -0.14 | 3.75 | 0.37 | 9 | 1 | 46.32 |
| BiSI | Pnam (62) | 1.18 | 1.32 | -0.69 | 0.05 | -0.64 | 4.79 | 0.53 | 2 | 1 | 29.59 |
| BiSeI | Pnma (62) | 0.91 | 1.03 | -0.68 | 0.04 | -0.65 | 5.89 | 0.25 | 2 | 1 | 26.83 |
| BiSBr | Pnam (62) | 1.32 | 1.35 | -0.53 | -0.11 | -0.64 | 6.21 | 0.24 | 2 | 1 | 30.10 |
| SbSI | P2$_1$2$_1$2$_1$ (19) | 1.28 | 1.46 | -0.23 | 0.11 | -0.11 | 2.84 | 0.91 | 2 | 2 | 31.59 |
| SbSI | Pna2$_1$ (33) | 1.45 | 1.60 | -0.17 | 0.11 | -0.06 | 2.06 | 1.31 | 2 | 2 | 69.72 |
| SbSI | Pnam (62) | 1.45 | 1.60 | -0.17 | 0.11 | -0.06 | 2.06 | 1.25 | 4 | 2 | 69.38 |
| SbSeI | Pnma (62) | 1.16 | 1.29 | -0.20 | 0.10 | -0.10 | 4.37 | 0.59 | 2 | 2 | 43.94 |
| RbBiS$_2$ | R$\bar{3}$mH (166) | 1.12 | 1.47 | -0.22 | -0.04 | -0.27 | 10.96 | 0.20 | 1 | 2 | 37.94 |
| InAlI$_4$ | P12$_1$/m1 (11) | 2.87 | 2.87 | -0.15 | 0.12 | -0.03 | 1.49 | 3.90 | 2 | 2 | 19.59 |
| CaInBr$_3$ | Cmcm (63) | 3.12 | 3.28 | -0.04 | 0.02 | -0.02 | 1.43 | 0.27 | 1 | 2 | 69.26 |
| KSnSb | P6$_3$mc (186) | 0.18 | 0.38 | -0.08 | 0.12 | 0.04 | 0.25 | 0.04 | 0.5 | 4 | 2.92 |
| Cs$_3$Sb$_2$I$_9$ | P$\bar{3}$m1 (164) | 1.40 | 1.41 | -0.14 | 0.20 | 0.05 | 2.19 | 0.25 | 1 | 7 | 13.06 |
| Cs$_3$Sb$_2$I$_9$ | P6$_3$/mmc (194) | 1.79 | 1.91 | -0.10 | 0.16 | 0.06 | 3.73 | 0.50 | 2 | 2 | 9.68 |

| | | | | | | | | | | |
|---|---|---|---|---|---|---|---|---|---|---|
| Cs$_3$Bi$_2$I$_9$ | P6$_3$/mmc (194) | 1.90 | 2.04 | -0.46 | 0.24 | -0.22 | 4.63 | 1.79 | 2 | 3 | 9.63 |
| (MA)GeI$_3$ | Pm$\bar{3}$m (221) | 0.82 | 0.82 | -0.18 | 0.11 | -0.06 | 0.43 | 0.51 | 2 | 2 | 6.89 |
| (MA)SnI$_3$ | Pm$\bar{3}$m (221) | 0.12 | 0.12 | -0.29 | 0.05 | -0.24 | 0.12 | 0.65 | 1 | 1 | 25.57 |
| (MA)PbI$_3$ | Pm$\bar{3}$m (221) | 0.31 | 0.31 | -1.12 | 0.13 | -1.00 | 0.10 | 0.16 | 1 | 1 | 20.07 |

It is important to draw attention to several conclusions from this table. Firstly, many of these materials are relatively ionic, like MAPbI$_3$, making it difficult to achieve smaller bandgaps (note again that the bandgaps presented here are underestimated, and many will be >2 eV in reality). Secondly, anisotropic crystal structures with stereochemically active lone pairs often lead to higher hole effective masses. Thirdly, most of the materials presented here demonstrate large ionic dielectric constants, independent of whether they have a polar space group. It is clear that MAPbI$_3$ offers an unusual mix of ionicity, high dielectric constant, low effective masses, and a lower bandgap for PV; few semiconductors can claim excellent performance across all categories.

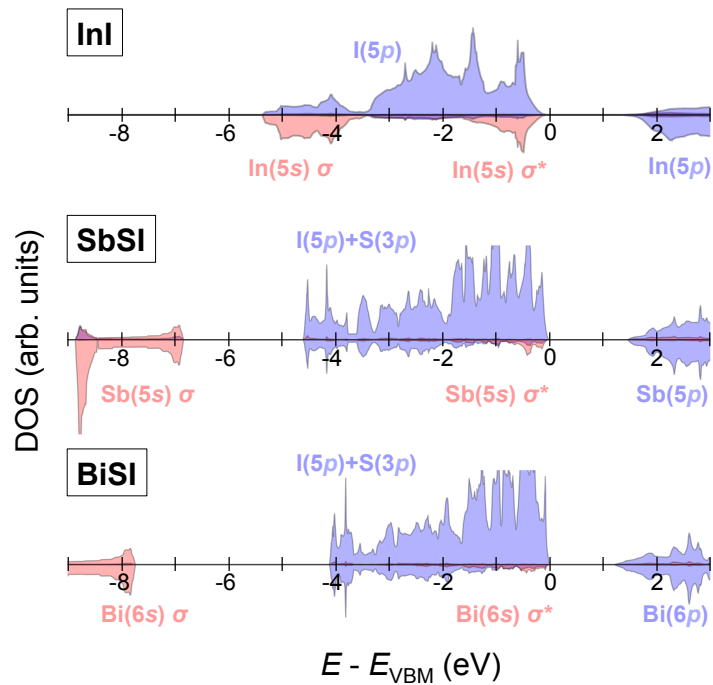

Fig. 4: Density of states plotted for three materials identified here, including (a) InI, (b) SbSI, and (c) BiSI. The partial densities of states are separated into their atomic orbital contributions, with anions on the positive $y$-axis and cations on the negative axis. The cation $s$ states separate into bonding $\sigma$ (lower energy) and antibonding $\sigma^*$ (higher energy) orbitals.

Another conclusion consistent with prior observations is that compounds formed with 6$p$ block cations show a significantly larger contribution of spin-orbit coupling in the conduction band, *vs*. those formed from the 5$p$ block. While Pb and Tl compounds are not included here, the Bi compounds show a much stronger spin-orbit contribution than do the In, Sn, and Sb compounds. This may suggest that the Bi-containing compounds are most suitable for achieving similarly shallow anion vacancy levels as in MAPbI$_3$.

To illustrate the similarities in band-structure to MAPbI$_3$, the band orbital diagrams and density of states are presented in Fig. 4 for a handful of materials. As anticipated, they all demonstrate the same cation antibonding *s*-orbital contributions at the valence band maximum.

## VI. SUMMARY

In this perspective piece, we have proposed new search criteria for the discovery of semiconductors for optoelectronic applications. This strategy builds upon the recent success of methyl-ammonium lead iodide as a PV material, which achieves excellent device efficiencies in the presence of defects.

While calculating defect-limited minority carrier mobility and lifetime is challenging, we propose that more fundamental electronic structure properties are indicative of defect tolerance, demonstrated by models for defect-assisted recombination and scattering. These models are consistent with emerging experimental and theoretical observations of MAPbX$_3$, as well as other semiconductor systems, which share defect tolerance. Most importantly, MAPbX$_3$ is not unique in these properties, and we identify a broad class of semiconductors containing partially oxidized cations, as well as several specific instances that may share these properties. To demonstrate this, we search through the MaterialsProject.org database of materials, and establish a set of criteria for future identification of promising defect-tolerant semiconductors.

Many fields would benefit from the development of semiconductors of a wide range of bandgaps that can achieve low non-radiative recombination rates and good transport – green LEDs, thin-film PV, and photodetectors, to name a few. We hope that this framework will provide a new direction for materials discovery in these fields, and in doing so, bring under-studied semiconductors to the fore to help build the next generation of energy-generating, light-emitting, sensing, and computing devices.

## METHODS

All density functional theory (DFT) and related calculations are performed using VASP computer code.[90] The GGA-PBE functional form of the exchange-correlation functional[91] has been employed in this work, together with the projector-augmented wave (PAW) formalism.[92] Full atomic, volume and cell shape relaxations are performed at the GGA-PBE level using the numerical setup, including *k*-point sampling and various cutoffs, following Ref. [93]. For the various electronic structure properties reported in Table 1, such as bandgaps, spin-orbit interaction contributions to band edges, DOS effective masses and band degeneracies; the denser *k*-point grid is used. For compatibility between different calculations, the number of *k*-points per atom was kept fixed at *Natoms*Nkpts*=6000, which leads to the convergence of relevant electronic structure properties.[55] The spin-orbit (SO) interaction is included at the level of first order perturbation to the DFT results. Ionic dielectric tensors are computed from the interatomic force constants obtained using density functional perturbation theory.[94] The dielectric constants reported in Table 1 are computed as one third of the trace ($Tr(\varepsilon)/3$) of the dielectric tensors.

**ACKNOWLEDGMENTS**

This work was supported as part of the Center for Next Generation Materials by Design (CMGMD), an Energy Frontier Research Center funded by the U.S. Department of Energy, Office of Science, Basic Energy Sciences. The authors thank R. Jaramillo, R. Chakraborty, V. Steinmann, and R. Kurchin (MIT) as well as S. Lany and A. Zakutayev (NREL) for helpful conversations.

**REFERENCES**

[1] W. Shockley, H.J. Queisser, Detailed Balance Limit of Efficiency of p-n Junction Solar Cells, J. Appl. Phys. 32 (1961) 510–519. doi:10.1063/1.1736034.
[2] M.B. Prince, Silicon Solar Energy Converters, J. Appl. Phys. 26 (1955) 534–540. doi:10.1063/1.1722034.
[3] L. Yu, A. Zunger, Identification of Potential Photovoltaic Absorbers Based on First-Principles Spectroscopic Screening of Materials, Phys. Rev. Lett. 108 (2012) 068701. doi:10.1103/PhysRevLett.108.068701.
[4] L. Yu, R.S. Kokenyesi, D.A. Keszler, A. Zunger, Inverse Design of High Absorption Thin-Film Photovoltaic Materials, Adv. Energy Mater. 3 (2013) 43–48. doi:10.1002/aenm.201200538.
[5] C. Wadia, A.P. Alivisatos, D.M. Kammen, Materials Availability Expands the Opportunity for Large-Scale Photovoltaics Deployment, Environ. Sci. Technol. 43 (2009) 2072–2077. doi:10.1021/es8019534.
[6] D.M. Powell, M.T. Winkler, H.J. Choi, C.B. Simmons, D.B. Needleman, T. Buonassisi, Crystalline silicon photovoltaics: a cost analysis framework for determining technology pathways to reach baseload electricity costs, Energy Environ. Sci. 5 (2012) 5874–5883. doi:10.1039/C2EE03489A.
[7] T. Surek, Crystal growth and materials research in photovoltaics: progress and challenges, J. Cryst. Growth. 275 (2005) 292–304. doi:10.1016/j.jcrysgro.2004.10.093.
[8] I.L. Repins, H. Moutinho, S.G. Choi, A. Kanevce, D. Kuciauskas, P. Dippo, et al., Indications of short minority-carrier lifetime in kesterite solar cells, J. Appl. Phys. 114 (2013) 084507. doi:10.1063/1.4819849.
[9] N.M. Mangan, R.E. Brandt, V. Steinmann, R. Jaramillo, J.V. Li, J.R. Poindexter, et al., A path to 10% efficiency for tin sulfide devices, in: Photovolt. Spec. Conf. PVSC 2014 IEEE 40th, 2014: pp. 2373–2378. doi:10.1109/PVSC.2014.6925404.
[10] T. Unold, H.W. Schock, Nonconventional (Non-Silicon-Based) Photovoltaic Materials, Annu. Rev. Mater. Res. 41 (2011) 297–321. doi:10.1146/annurev-matsci-062910-100437.
[11] S.D. Stranks, G.E. Eperon, G. Grancini, C. Menelaou, M.J.P. Alcocer, T. Leijtens, et al., Electron-Hole Diffusion Lengths Exceeding 1 Micrometer in an Organometal Trihalide Perovskite Absorber, Science. 342 (2013) 341–344. doi:10.1126/science.1243982.
[12] Q. Dong, Y. Fang, Y. Shao, P. Mulligan, J. Qiu, L. Cao, et al., Electron-hole diffusion lengths >175 m in solution grown CH3NH3PbI3 single crystals, Science. 347 (2015) 967–970. doi:10.1126/science.aaa5760.

[13] N.J. Jeon, J.H. Noh, W.S. Yang, Y.C. Kim, S. Ryu, J. Seo, et al., Compositional engineering of perovskite materials for high-performance solar cells, Nature. 517 (2015) 476–480. doi:10.1038/nature14133.
[14] National Renewable Energy Laboratory, Best Research- Cell Efficiencies, Best Res. Cell Effic. (n.d.). www.nrel.gov/ncpv/images/ efficiency_chart.jpg.
[15] A. Kojima, K. Teshima, Y. Shirai, T. Miyasaka, Organometal Halide Perovskites as Visible-Light Sensitizers for Photovoltaic Cells, J. Am. Chem. Soc. 131 (2009) 6050–6051. doi:10.1021/ja809598r.
[16] M.A. Green, A. Ho-Baillie, H.J. Snaith, The emergence of perovskite solar cells, Nat. Photonics. 8 (2014) 506–514. doi:10.1038/nphoton.2014.134.
[17] C.C. Stoumpos, C.D. Malliakas, M.G. Kanatzidis, Semiconducting Tin and Lead Iodide Perovskites with Organic Cations: Phase Transitions, High Mobilities, and Near-Infrared Photoluminescent Properties, Inorg. Chem. 52 (2013) 9019–9038. doi:10.1021/ic401215x.
[18] J. Mattheis, J. Werner, U. Rau, Finite mobility effects on the radiative efficiency limit of pn-junction solar cells, Phys. Rev. B. 77 (2008) 085203. doi:10.1103/PhysRevB.77.085203.
[19] J.R. Davis, Jr., A. Rohatgi, R.H. Hopkins, P.D. Blais, P. Rai-Choudhury, J.R. McCormick, et al., Impurities in silicon solar cells, IEEE Trans. Electron Devices. 27 (1980) 677–687. doi:10.1109/T-ED.1980.19922.
[20] H.R. Moutinho, R.G. Dhere, M.M. Al-Jassim, C. Ballif, D.H. Levi, A.B. Swartzlander, et al., Study of CdTe/CdS solar cells using CSS CdTe deposited at low temperature, in: Photovolt. Spec. Conf. PVSC 2000 IEEE 28th, 2000: pp. 646–649.
[21] L. Kranz, C. Gretener, J. Perrenoud, D. Jaeger, S.S.A. Gerstl, R. Schmitt, et al., Tailoring Impurity Distribution in Polycrystalline CdTe Solar Cells for Enhanced Minority Carrier Lifetime, Adv. Energy Mater. 4 (2014) 1301400. doi:10.1002/aenm.201301400.
[22] I.L. Repins, W.K. Metzger, C.L. Perkins, J.V. Li, M.A. Contreras, Measured minority-carrier lifetime and CIGS device performance, in: Photovolt. Spec. Conf. PVSC 2009 IEEE 34th, 2009: pp. 000978–000983. doi:10.1109/PVSC.2009.5411126.
[23] O.D. Miller, E. Yablonovitch, S.R. Kurtz, Strong Internal and External Luminescence as Solar Cells Approach the Shockley #x2013;Queisser Limit, IEEE J. Photovolt. 2 (2012) 303–311. doi:10.1109/JPHOTOV.2012.2198434.
[24] E.A. Schiff, Hole mobilities and the physics of amorphous silicon solar cells, J. Non-Cryst. Solids. 352 (2006) 1087–1092. doi:10.1016/j.jnoncrysol.2005.11.074.
[25] W.K. Metzger, D. Albin, D. Levi, P. Sheldon, X. Li, B.M. Keyes, et al., Time-resolved photoluminescence studies of CdTe solar cells, J. Appl. Phys. 94 (2003) 3549–3555. doi:10.1063/1.1597974.
[26] D. Walter, F. Rosenits, F. Kopp, S. Reber, B. Berger, W. Warta, Determining the Minority Carrier Lifetime in Epitaxial Silicon Layers by Micro-Wave-Detected Photoconductivity Measurements, in: 25th Eur. Photovolt. Sol. Energy Conf., Valencia, Spain, 2010: pp. 2078 – 2083. doi:10.4229/25thEUPVSEC2010-2CV.3.1.
[27] T. Tiedje, C.R. Wronski, B. Abeles, J.M. Cebulka, Electron transport in hydrogenated amorphous silicon: drift mobility and junction capacitance, Sol. Cells. 2 (1980) 301–318. doi:10.1016/0379-6787(80)90034-4.
[28] R.A. Sinton, A. Cuevas, M. Stuckings, Quasi-steady-state photoconductance, a new method for solar cell material and device characterization, in: 25th Photovolt. Spec. Conf., IEEE, Washington, D.C., 1996: pp. 457–460.
[29] J.K. Katahara, H.W. Hillhouse, Quasi-Fermi level splitting and sub-bandgap absorptivity from semiconductor photoluminescence, J. Appl. Phys. 116 (2014) 173504. doi:10.1063/1.4898346.
[30] A. Jain, S.P. Ong, G. Hautier, W. Chen, W.D. Richards, S. Dacek, et al., Commentary: The Materials Project: A materials genome approach to accelerating materials innovation, APL Mater. 1 (2013) 011002. doi:10.1063/1.4812323.
[31] W.-J. Yin, T. Shi, Y. Yan, Unusual defect physics in CH3NH3PbI3 perovskite solar cell absorber, Appl. Phys. Lett. 104 (2014) 063903. doi:10.1063/1.4864778.
[32] W.-J. Yin, T. Shi, Y. Yan, Unique Properties of Halide Perovskites as Possible Origins of the Superior Solar Cell Performance, Adv. Mater. 26 (2014) 4653–4658. doi:10.1002/adma.201306281.
[33] Y. Yamada, T. Nakamura, M. Endo, A. Wakamiya, Y. Kanemitsu, Near-band-edge optical responses of solution-processed organic?inorganic hybrid perovskite CH3NH3PbI3 on mesoporous TiO2 electrodes, Appl. Phys. Express. 7 (2014) 032302. doi:10.7567/APEX.7.032302.

[34] A. Walsh, Principles of Chemical Bonding and Band Gap Engineering in Hybrid Organic-Inorganic Halide Perovskites, J. Phys. Chem. C. 119 (2015) 5755–5760. doi:10.1021/jp512420b.

[35] M.H. Du, Efficient carrier transport in halide perovskites: theoretical perspectives, J. Mater. Chem. A. 2 (2014) 9091–9098. doi:10.1039/C4TA01198H.

[36] S.B. Zhang, S.-H. Wei, A. Zunger, H. Katayama-Yoshida, Defect physics of the CuInSe$_2$ chalcopyrite semiconductor, Phys. Rev. B. 57 (1998) 9642.

[37] A. Zakutayev, C.M. Caskey, A.N. Fioretti, D.S. Ginley, J. Vidal, V. Stevanović, et al., Defect Tolerant Semiconductors for Solar Energy Conversion, J. Phys. Chem. Lett. 5 (2014) 1117–1125. doi:10.1021/jz5001787.

[38] D. Shi, V. Adinolfi, R. Comin, M. Yuan, E. Alarousu, A. Buin, et al., Low trap-state density and long carrier diffusion in organolead trihalide perovskite single crystals, Science. 347 (2015) 519–522. doi:10.1126/science.aaa2725.

[39] K. Tvingstedt, O. Malinkiewicz, A. Baumann, C. Deibel, H.J. Snaith, V. Dyakonov, et al., Radiative efficiency of lead iodide based perovskite solar cells, Sci. Rep. 4 (2014). doi:10.1038/srep06071.

[40] C. Wehrenfennig, G.E. Eperon, M.B. Johnston, H.J. Snaith, L.M. Herz, High Charge Carrier Mobilities and Lifetimes in Organolead Trihalide Perovskites, Adv. Mater. 26 (2014) 1584–1589. doi:10.1002/adma.201305172.

[41] T. Umebayashi, K. Asai, T. Kondo, A. Nakao, Electronic structures of lead iodide based low-dimensional crystals, Phys. Rev. B. 67 (2003) 155405. doi:10.1103/PhysRevB.67.155405.

[42] E.M. Miller, Y. Zhao, C.C. Mercado, S.K. Saha, J.M. Luther, K. Zhu, et al., Substrate-controlled band positions in $CH_3NH_3PbI_3$ perovskite films, Phys Chem Chem Phys. 16 (2014) 22122–22130. doi:10.1039/C4CP03533J.

[43] P. Umari, E. Mosconi, F. De Angelis, Relativistic Solar Cells, Sci. Rep. 4 (2014) 4467.

[44] S.M. Sze, K.N. Kwok, Physics of Semiconductor Devices, 3rd ed., John Wiley & Sons, Hoboken, NJ, 2007.

[45] J.M. Frost, K.T. Butler, F. Brivio, C.H. Hendon, M. van Schilfgaarde, A. Walsh, Atomistic origins of high-performance in hybrid halide perovskite solar cells, Nano Lett. 14 (2014) 2584–2590.

[46] R.H. Bube, Photoelectronic Properties of Semiconductors, Cambridge University Press, 1992.

[47] Q. Lin, A. Armin, R.C.R. Nagiri, P.L. Burn, P. Meredith, Electro-optics of perovskite solar cells, Nat. Photonics. 9 (2015) 106–112. doi:10.1038/nphoton.2014.284.

[48] E.J. Juarez-Perez, R.S. Sanchez, L. Badia, G. Garcia-Belmonte, Y.S. Kang, I. Mora-Sero, et al., Photoinduced Giant Dielectric Constant in Lead Halide Perovskite Solar Cells, J. Phys. Chem. Lett. 5 (2014) 2390–2394. doi:10.1021/jz5011169.

[49] F. Brivio, A.B. Walker, A. Walsh, Structural and electronic properties of hybrid perovskites for high-efficiency thin-film photovoltaics from first-principles, APL Mater. 1 (2013) 042111. doi:10.1063/1.4824147.

[50] J. Kim, S.-H. Lee, J.H. Lee, K.-H. Hong, The Role of Intrinsic Defects in Methylammonium Lead Iodide Perovskite, J. Phys. Chem. Lett. 5 (2014) 1312–1317. doi:10.1021/jz500370k.

[51] S.D. Lester, F.A. Ponce, M.G. Craford, D.A. Steigerwald, High dislocation densities in high efficiency GaN-based light-emitting diodes, Appl. Phys. Lett. 66 (1995) 1249. doi:10.1063/1.113252.

[52] M. Carmody, S. Mallick, J. Margetis, R. Kodama, T. Biegala, D. Xu, et al., Single-crystal II-VI on Si single-junction and tandem solar cells, Appl. Phys. Lett. 96 (2010) 153502. doi:10.1063/1.3386529.

[53] Y.S. Lee, M.T. Winkler, S.C. Siah, R. Brandt, T. Buonassisi, Hall mobility of cuprous oxide thin films deposited by reactive direct-current magnetron sputtering, Appl. Phys. Lett. 98 (2011) 192115. doi:10.1063/1.3589810.

[54] J.Y.W. Seto, The electrical properties of polycrystalline silicon films, J. Appl. Phys. 46 (1975) 5247. doi:10.1063/1.321593.

[55] J. Yan, P. Gorai, B. Ortiz, S. Miller, S.A. Barnett, T. Mason, et al., Material descriptors for predicting thermoelectric performance, Energy Env. Sci. 8 (2014) 983–994. doi:10.1039/C4EE03157A.

[56] J. Feng, Mechanical properties of hybrid organic-inorganic $CH_3NH_3BX_3$ (B = Sn, Pb; X = Br, I) perovskites for solar cell absorbers, APL Mater. 2 (2014) 081801. doi:10.1063/1.4885256.

[57] J.J. Krich, B.I. Halperin, A. Aspuru-Guzik, Nonradiative lifetimes in intermediate band photovoltaics—Absence of lifetime recovery, J. Appl. Phys. 112 (2012) 013707. doi:10.1063/1.4732085.

[58] S. De Wolf, J. Holovsky, S.-J. Moon, P. Löper, B. Niesen, M. Ledinsky, et al., Organometallic Halide Perovskites: Sharp Optical Absorption Edge and Its Relation to Photovoltaic Performance, J. Phys. Chem. Lett. 5 (2014) 1035–1039. doi:10.1021/jz500279b.
[59] P.J. Sellin, Thick film compound semiconductors for X-ray imaging applications, Nucl. Instrum. Methods Phys. Res. Sect. A. 563 (2006) 1–8. doi:10.1016/j.nima.2006.01.110.
[60] G. Hautier, A. Miglio, G. Ceder, G.-M. Rignanese, X. Gonze, Identification and design principles of low hole effective mass p-type transparent conducting oxides, Nat. Commun. 4 (2013) 2292. doi:doi:10.1038/ncomms3292.
[61] R. Nechache, C. Harnagea, S. Licoccia, E. Traversa, A. Ruediger, A. Pignolet, et al., Photovoltaic properties of Bi2FeCrO6 epitaxial thin films, Appl. Phys. Lett. 98 (2011) 202902. doi:10.1063/1.3590270.
[62] J. Rödel, W. Jo, K.T.P. Seifert, E.-M. Anton, T. Granzow, D. Damjanovic, Perspective on the Development of Lead-free Piezoceramics, J. Am. Ceram. Soc. 92 (2009) 1153–1177. doi:10.1111/j.1551-2916.2009.03061.x.
[63] A.-V. Mudring, Thallium Halides – New Aspects of the Stereochemical Activity of Electron Lone Pairs of Heavier Main-Group Elements, Eur. J. Inorg. Chem. 2007 (2007) 882–890. doi:10.1002/ejic.200600975.
[64] M.-H. Du, D.J. Singh, Enhanced Born charge and proximity to ferroelectricity in thallium halides, Phys. Rev. B. 81 (2010) 144114.
[65] S.P. Ong, W.D. Richards, A. Jain, G. Hautier, M. Kocher, S. Cholia, et al., Python Materials Genomics (pymatgen): A robust, open-source python library for materials analysis, Comput. Mater. Sci. 68 (2013) 314–319. doi:10.1016/j.commatsci.2012.10.028.
[66] A. Walsh, D.J. Payne, R.G. Egdell, G.W. Watson, Stereochemistry of post-transition metal oxides: revision of the classical lone pair model, Chem. Soc. Rev. 40 (2011) 4455. doi:10.1039/c1cs15098g.
[67] V. Stevanović, K. Hartman, R. Jaramillo, S. Ramanathan, T. Buonassisi, P. Graf, Variations of ionization potential and electron affinity as a function of surface orientation: The case of orthorhombic SnS, Appl. Phys. Lett. 104 (2014) 211603. doi:doi:10.1063/1.4879558.
[68] I. Chung, J.-H. Song, J. Im, J. Androulakis, C.D. Malliakas, H. Li, et al., CsSnI3: Semiconductor or metal? High electrical conductivity and strong near-infrared photoluminescence from a single material. High hole mobility and phase-transitions, J. Am. Chem. Soc. 134 (2012) 8579–8587. doi:10.1021/ja301539s.
[69] R. Nechache, C. Harnagea, S. Li, L. Cardenas, W. Huang, J. Chakrabartty, et al., Bandgap tuning of multiferroic oxide solar cells, Nat. Photonics. 9 (2015) 61–67. doi:10.1038/nphoton.2014.255.
[70] A.T. Lintereur, W. Qiu, J.C. Nino, J. Baciak, Characterization of bismuth tri-iodide single crystals for wide band-gap semiconductor radiation detectors, Nucl. Instrum. Methods Phys. Res. Sect. A. 652 (2011) 166–169. doi:10.1016/j.nima.2010.12.013.
[71] J. Vidal, S. Lany, J. Francis, R. Kokenyesi, J. Tate, Structural and electronic modification of photovoltaic SnS by alloying, J. Appl. Phys. 115 (2014) 113507. doi:10.1063/1.4868974.
[72] E.A. Secco, A. Sharma, Structure stabilization: Locking-in fast cation conductivity phase in TlI, J. Phys. Chem. Solids. 56 (1995) 251–254. doi:10.1016/0022-3697(94)00172-3.
[73] R. Nitsche, W.J. Merz, Photoconduction in ternary V-VI-VII compounds, J. Phys. Chem. Solids. 13 (1960) 154–155.
[74] N.T. Hahn, A.J.E. Rettie, S.K. Beal, R.R. Fullon, C.B. Mullins, n-BiSI Thin Films: Selenium Doping and Solar Cell Behavior, J. Phys. Chem. C. 116 (2012) 24878–24886. doi:10.1021/jp3088397.
[75] N.T. Hahn, J.L. Self, C.B. Mullins, BiSI Micro-Rod Thin Films: Efficient Solar Absorber Electrodes?, J. Phys. Chem. Lett. 3 (2012) 1571–1576. doi:10.1021/jz300515p.
[76] X. Zhang, L. Zhang, T. Xie, D. Wang, Low-Temperature Synthesis and High Visible-Light-Induced Photocatalytic Activity of BiOI/TiO$_2$ Heterostructures, J. Phys. Chem. C. 113 (2009) 7371–7378. doi:10.1021/jp900812d.
[77] K. Zhao, X. Zhang, L. Zhang, The first BiOI-based solar cells, Electrochem. Commun. 11 (2009) 612–615. doi:10.1016/j.elecom.2008.12.041.
[78] K.T. Butler, J.M. Frost, A. Walsh, Ferroelectric materials for solar energy conversion: photoferroics revisited, Energy Environ. Sci. 8 (2015) 838–848. doi:10.1039/C4EE03523B.
[79] B.V. Gabrel'yan, A.A. Lavrentiev, I.Y. Nikiforov, V.V. Sobolev, Electronic energy structure of MBiS2 (M = Li, Na, K) calculated with allowance for the difference between the M-S and Bi-S bond lengths, J. Struct. Chem. 49 (2008) 788–794. doi:10.1007/s10947-008-0140-2.

[80] S. Kang, Y. Hong, Y. Jeon, A Facile Synthesis and Characterization of Sodium Bismuth Sulfide (NaBiS$_2$) under Hydrothermal Condition, Bull. Korean Chem. Soc. 35 (2014) 1887–1890. doi:10.5012/bkcs.2014.35.6.1887.

[81] T.J. McCarthy, S.P. Ngeyi, J.H. Liao, D.C. DeGroot, T. Hogan, C.R. Kannewurf, et al., Molten salt synthesis and properties of three new solid-state ternary bismuth chalcogenides, .beta.-CsBiS2, .gamma.-CsBiS2, and K2Bi8Se13, Chem. Mater. 5 (1993) 331–340. doi:10.1021/cm00027a016.

[82] T. Timofte, A.-V. Mudring, Indium(I) Tetraiodoaluminate, InAlI4, Z. Für Anorg. Allg. Chem. 634 (2008) 622–623. doi:10.1002/zaac.200700525.

[83] S.V. Mel'nikova, A.I. Zaitsev, Ferroelectric phase transition in Cs3Bi2I9, Phys. Solid State. 39 (1997) 1652–1654.

[84] E.Y. Peresh, V.I. Sidei, N.I. Gaborets, O.V. Zubaka, I.P. Stercho, I.E. Barchii, Influence of the average atomic number of the A2TeC6 and A3B2C9 (A = K, Rb, Cs, Tl(I); B = Sb, Bi; C = Br, I) compounds on their melting point and band gap, Inorg. Mater. 50 (2014) 101–106. doi:10.1134/S0020168514010166.

[85] R. Jakubas, L. Sobczyk, Phase transitions in alkylammonium halogenoantimonates and bismuthates, Phase Transit. 20 (1990) 163–193. doi:10.1080/01411599008206873.

[86] L.-M. Wu, X.-T. Wu, L. Chen, Structural overview and structure–property relationships of iodoplumbate and iodobismuthate, Coord. Chem. Rev. 253 (2009) 2787–2804. doi:10.1016/j.ccr.2009.08.003.

[87] T. Plackowski, D. Wlosewicz, P.E. Tomaszewski, J. Baran, M.K. Marchewka, Specific heat of (NH2(CH3)2)3Bi2I9, Acta Phys. Pol. A. 87 (1995) 635–641.

[88] G. Kieslich, S. Sun, A.K. Cheetham, Solid-state principles applied to organic–inorganic perovskites: new tricks for an old dog, Chem Sci. 5 (2014) 4712–4715. doi:10.1039/C4SC02211D.

[89] M.R. Filip, G.E. Eperon, H.J. Snaith, F. Giustino, Steric engineering of metal-halide perovskites with tunable optical band gaps, Nat. Commun. 5 (2014) 5757. doi:10.1038/ncomms6757.

[90] G. Kresse, J. Furthmüller, Efficient iterative schemes for \textit{ab initio} total-energy calculations using a plane-wave basis set, Phys. Rev. B. 54 (1996) 11169–11186. doi:10.1103/PhysRevB.54.11169.

[91] J.P. Perdew, K. Burke, M. Ernzerhof, Generalized Gradient Approximation Made Simple, Phys. Rev. Lett. 77 (1996) 3865–3868. doi:10.1103/PhysRevLett.77.3865.

[92] P.E. Blöchl, Projector augmented-wave method, Phys. Rev. B. 50 (1994) 17953–17979. doi:10.1103/PhysRevB.50.17953.

[93] V. Stevanović, S. Lany, X. Zhang, A. Zunger, Correcting density functional theory for accurate predictions of compound enthalpies of formation: Fitted elemental-phase reference energies, Phys. Rev. B. 85 (2012) 115104. doi:10.1103/PhysRevB.85.115104.

[94] X. Wu, D. Vanderbilt, D.R. Hamann, Systematic treatment of displacements, strains, and electric fields in density-functional perturbation theory, Phys. Rev. B. 72 (2005) 035105. doi:10.1103/PhysRevB.72.035105.